\documentclass[twocolumn,showpacs,preprintnumbers,amsmath,amssymb,nofootinbib]{revtex4}
\usepackage{graphicx}
\usepackage{dcolumn}
\usepackage{bm,color}

\newcommand{\refeq}[1]{~(\ref{#1})}

\newcommand{\eqd}{\stackrel{d}{=}}
\newcommand{\pslash}{p\!\!\!\!\diagup}
\newcommand{\dslash}{\partial\!\!\!\!\diagup}

\begin{document}

\title{ \textbf{Mass spectrum and L\évy--Schr\"odinger relativistic equation}}
\author{Nicola \surname{Cufaro Petroni}}
 \email{cufaro@ba.infn.it}
 \affiliation{Department of Mathematics and TIRES, Bari
 University;\\
 INFN Sezione di Bari, \\
via E Orabona 4, 70125 Bari, Italy}
\author{Modesto Pusterla}
 \email{pusterla@pd.infn.it}
 \affiliation{Department of Physics, Padova
 University;\\
 INFN Sezione di Padova, \\
via F.\ Marzolo 8, 35100 Padova, Italy}

\begin{abstract}
\noindent We introduce a modification in the relativistic equations in such a way that (1) the
relativistic Schr\"odinger equations can always be based on an underlying L\évy process, (2)
several families of particles with different rest masses can be selected, and finally (3) the
corresponding Feynman diagrams are convergent when we have at least three different masses.
\end{abstract}

\pacs{03.65.Pm, 02.50.Ey, 12.38.Bx}

\maketitle

\section{Introduction and notations}\label{intro}

In this note we adopt the space-time relativistic approach of Feynman's propagators (for bosons
and fermions) instead of the canonical Lagrangian-Hamiltonian quantized field theory. Indeed the
former alternative is preferred to the latter for the developments of our basic ideas that exhibit
the connection between the propagator of quantum mechanics and L\évy's stochasticity. More
precisely the relativistic Feynman propagators are here linked with a dynamical theory based on a
particular L\évy stochastic process. This point, already mentioned in a previous
paper~\cite{cufaro09}, is here analyzed thoroughly with the purpose of deducing its consequences
for the case of fundamental fermions and bosons (quarks, leptons, gluons etc$\ldots$) of the
Standard Model (SM) characterized by the symmetry $SU_C(3)\times SU_L(2)\times U(1)$. To this end
we now recall that a L\évy process is a stochastic process $X(t),\;t\geq0$ on a probability space
$(\Omega,\mathcal{F},\mathbb{P})$ such that
\begin{itemize}
    \item $X(0)=0,\quad \mathbb{P}$-q.o.
    \item $X(t)$ has independent and stationary increments: for each $n$ and for very choice of
    $0\leq t_1<t_2<\ldots<t_n<+\infty$ the increments $X(t_{j+1})-X(t_j)$ are independent and $X(t_{j+1})-X(t_j)\eqd
    X(t_{j+1}-X(t_j)$;
    \item $X(t)$ is stochastically continuous: for every $a>0$ and for every $s$
    \begin{equation*}
        \lim_{t\to s}\mathbb{P}\left(|X(t)-X(s)|>a\right)=0.
    \end{equation*}
\end{itemize}
To simplify the notation we will consider in the following one--dimensional (the $n$--dimensional
extension would not be a difficult task) L\évy processes: it is well
known~\cite{sato,applebaum,cufaro08} that all its laws are infinitely divisible, but we will be
mainly interested in the non stable (and in particular non Gaussian) case. In other words the
characteristic function of the process $\Delta t$--increment is $[\varphi(u)]^{\Delta t/\tau}$
where $\varphi$ is infinitely divisible, but not stable~\footnote{A law $\varphi$ is said to be
infinitely divisible if for every $n$ it exists a characteristic function $\varphi_n$ such that
$\varphi=\varphi_n^n$; on the other hand it is said to be stable when for every $c>0$ it is always
possible to find $a>0$ and $b\in\mathbf{R}$ such that $e^{ibu}\varphi(au)=[\varphi(u)]^c$. Every
stable law is also infinitely divisible.}, and $\tau$ is a time scale parameter. The transition
probability density $p(2|1)=p(x_2,t_2|x_1,t_1)$ of a particle moving from the space-time point 1
to 2 then is
\begin{equation}\label{transpdf}
    p(2|1)=\frac{1}{2\pi}\int_{-\infty}^{+\infty}du\,[\varphi(u)]^{(t_2-t_1)/\tau}e^{-iu(x_2-x_1)}
\end{equation}
In analogy with the non relativistic Wiener case we obtain for the motion of a free particle the
Feynman propagator $\mathcal{K}(2|1)=\mathcal{K}(x_2,t_2|x_1,t_1)$ as
\begin{equation}\label{propagator}
    \mathcal{K}(2|1)=\frac{1}{2\pi}\int_{-\infty}^{+\infty}du\,[\varphi(u)]^{i(t_2-t_1)/\tau}e^{-iu(x_2-x_1)}
\end{equation}
and the corresponding wave function evolution is
\begin{equation}\label{evolution}
    \psi(x,t)=\int_{-\infty}^{+\infty}dx'\,\mathcal{K}(x,t|x',t')\psi(x',t').
\end{equation}
From\refeq{propagator} and\refeq{evolution} we easily obtain~\cite{cufaro09}
\begin{equation*}
    i\partial_t\psi=-\frac{1}{\tau}\eta(\partial_x)\psi
\end{equation*}
where $\eta=\log\varphi$ and $\eta(\partial_x)$ is a pseudodifferential operator with symbol
$\eta(u)$ defined through the use of Fourier transforms~\cite{applebaum,cont,taylor,jacob}. It
plays the role of the generator of the semigroup $T_t=e^{t\eta(\partial_x)/\tau}$ operating on the
Banach space of the measurable, bounded functions~\cite{applebaum,cont,taylor,jacob}.

It is very well known~\cite{sato,applebaum}, on the other hand, that $\varphi$ represents an
infinitely divisible law if and only if $\eta(u)=\log\varphi(u)$ satisfies the L\évy--Khintchin
formula
\begin{equation}\label{LK}
    \eta(u)=i\gamma
    u-\frac{\beta^2u^2}{2}+\int_{\mathbb{R}}\left[e^{iux}-1-iux\,I_{[-1,1]}(x)\right]\,\nu(dx)
\end{equation}
where $\gamma,\beta\in\mathbb{R}$, $I_{[-1,1]}(x)$ is the indicator of $[-1,1]$, and $\nu(dx)$ is
the L\évy measure, namely a measure on $\mathbb{R}$ such that $\nu(\{0\})=0$ and
\begin{equation*}
    \int_{\mathbb{R}}(x^2\wedge1)\,\nu(dx)<+\infty.
\end{equation*}
In the case of a centered, symmetric law the equation\refeq{LK} simplifies in
\begin{equation}\label{LKsymm}
    \eta(u)=-\frac{\beta^2u^2}{2}+\int_{\mathbb{R}}(\cos ux-1)\,\nu(dx)
\end{equation}
and $\eta(u)$ becomes even and real. As a consequence the corresponding operator
$\eta(\partial_x)$ is self--adjoint and acts on propagators and wave functions according to the
L\évy--Schr\"odinger integro--differential equation
\begin{eqnarray}\label{lseq}
    i\partial_t\psi(x,t)&=&-\frac{1}{\tau}\,\eta(\partial_x)\psi(x,t)\nonumber\\
                             &=&-\frac{\beta^2}{2\tau}\,\partial^2_x\psi(x,t)\\
                        &&\qquad
                        -\frac{1}{\tau}\int_{\mathbb{R}}\left[\psi(x+y,t)-\psi(x,t)\right]\,\nu(dy).\nonumber
\end{eqnarray}
The integral term accounts for the jumps in the trajectories of the
underlying stochastic process, while an action $\alpha$ with
$\beta^2=\alpha\tau/m$ provides the usual differential term of the
Schr\"odinger equation. For $\beta=0$ a pure jump
L\évy--Schr\"odinger equation is obtained
\begin{equation}\label{symmlseq}
     i\partial_t\psi(x,t)=-\frac{1}{\tau}\int_{\mathbb{R}}\left[\psi(x+y,t)-\psi(x,t)\right]\,\nu(dy).
\end{equation}

\section{Stationary solutions for the free particle}

Equation\refeq{lseq} allows a simple stationary solution: if we consider
\begin{equation*}
    \psi(x,t)=e^{-iE_0t/\alpha}\phi(x),\qquad\quad\alpha=\frac{m\beta^2}{\tau}
\end{equation*}
we have then
\begin{equation}\label{statsol}
    E_0\phi(x)=-\frac{\alpha^2}{2m}\,\phi''(x)-\frac{\alpha}{\tau}\int_\mathbb{R}[\phi(x+y)-\phi(x)]\,\nu(dy),
\end{equation}
and for a plane wave $\phi(x)=e^{\pm iux}$ from\refeq{LKsymm} with a
symmetric L\évy noise
\begin{eqnarray*}
  E_0\phi(x) 
           &=&-\frac{\alpha}{\tau}\left[-\frac{\beta^2u^2}{2}+\int_\mathbb{R}\left(e^{\pm iuy}-1\right)\,\nu(dy)\right]e^{\pm iux}  \\
           &=&-\frac{\alpha}{\tau}\left[-\frac{\beta^2u^2}{2}+\int_\mathbb{R}(\cos uy-1)\,\nu(dy)\right]\phi(x)\\
           &=&-\frac{\alpha}{\tau}\eta(u)\phi(x)
\end{eqnarray*}
which is satisfied when $E_0=-\alpha\eta(u)/\tau$. Finally by taking
$p=\alpha u$ for the momentum we obtain the relevant equation
\begin{equation}\label{Ep}
    E_0=-\frac{\alpha}{\tau}\,\eta\left(\frac{p}{\alpha}\right).
\end{equation}

\section{Relativistic quantum mechanics}\label{rqm}

Equation\refeq{Ep} connects the kinetic energy of a forceless particle to the logarithmic
characteristic of a L\évy process: if we take in particular the non stable law $\eta(u)=
1-\sqrt{1+a^2u^2}$ with the following identification of the parameters
\begin{equation*}
    \alpha=\hbar,\qquad\quad\frac{\hbar}{\tau}=mc^2,\qquad\quad a=\frac{\hbar}{mc},\qquad\quad p=\hbar u.
\end{equation*}
we are led to the formula
\begin{equation}\label{relEp}
    E_0=-mc^2\eta\left(\frac{p}{\hbar}\right)=E-mc^2=\sqrt{m^2c^4+p^2c^2}-mc^2
\end{equation}
which is the well--known relativistic kinetic energy for a particle
of mass $m$. The Schr\"odinger equation of a relativistic
free--particle is easily obtained from\refeq{relEp} by
reinterpreting as usual $E$ and $p$ respectively as the operators
$i\hbar\partial_t$ and $-i\hbar\partial_x$:
\begin{equation}\label{releq}
    i\hbar\partial_t\psi(x,t)=\sqrt{m^2c^4-\hbar^2c^2\partial^2_x}\,\psi(x,t),
\end{equation}
but this comes also from\refeq{lseq} after having absorbed the mass
energy term $-mc^2$ of\refeq{relEp} into a phase factor
$e^{imc^2t/\hbar}$. In three dimensions\refeq{releq} reads
\begin{equation}\label{releq3}
    i\hbar\partial_t\psi(x,t)=\sqrt{m^2c^4-\hbar^2c^2\bm\nabla^2}\,\psi(x,t)
\end{equation}
It has been shown~\cite{applebaum,ichinose} that the L\évy process
behind the equations\refeq{releq} and\refeq{releq3} is a pure jump
process~\cite{applebaum,cufaro09} with an absolutely continuous
L\évy measure $\nu(dx)=W(x)dx$ and
\begin{equation}
    W(x)=\frac{1}{\pi|x|}\,K_1\left(\frac{|x|}{a}\right)=\frac{1}{\pi|x|}\,K_1\left(\frac{mc}{\hbar}\,|x|\right)
\end{equation}
($K_\nu$ are the modified Bessel functions~\cite{abramowitz}), that
in three dimensions becomes
\begin{equation}
    W(\bm x)=\frac{1}{2a\pi^2|\bm x|^2}\,K_2\left(\frac{|\bm x|}{a}\right)=\frac{mc}{2\hbar\pi^2|\bm x|^2}\,K_2\left(\frac{mc}{\hbar}\,|\bm x|\right)
\end{equation}
while from\refeq{symmlseq} the equation\refeq{releq} becomes equivalent to
\begin{eqnarray}\label{intdiff}
    \lefteqn{i\hbar\partial_t\psi(x,t)} \\
    &&=-mc^2\int_{\mathbb{R}}\frac{\psi(x+y,t)-\psi(x,t)}{\pi|y|}\,K_1\left(\frac{mc}{\hbar}\,|y|\right)dy\nonumber
\end{eqnarray}
and in three dimensions
\begin{eqnarray}\label{intdiff3}
    \lefteqn{i\hbar\partial_t\psi(\bm x,t)}\\
    &&=-mc^2\!\!\int_{\mathbb{R}^3}\!\!\frac{\psi(\bm x+\bm y,t)-\psi(\bm x,t)}{2\pi^2|\bm y|^2}\,\frac{mc}{\hbar}K_2\left(\frac{mc}{\hbar}\,|\bm y|\right)d^3\bm
    y\nonumber
\end{eqnarray}
From the equation\refeq{releq3} by the well known standard
procedures~\cite{bjorken} one derives (always for the free particle)
the Klein--Gordon and Dirac equations in three dimensions for the
wave functions and spinors $\psi$, respectively
\begin{eqnarray}
  \left(\square-\frac{m^2c^2}{\hbar^2}\right)\psi &=& 0, \label{KG}\\
  \left(i\gamma_\mu\partial^\mu-\frac{mc}{\hbar}\right)\psi  &=& 0.\label{D}
\end{eqnarray}
The Klein--Gordon and Dirac propagators verify instead the inhomogeneous equations (here
$\hbar=c=1$)
\begin{eqnarray}
  \left(\square_2-m^2\right)\mathcal{K}_{KG}(2|1) &=& \delta^4(2|1)\label{propKG} \\
  \left(i\gamma_\mu\partial_2^\mu-m\right)\mathcal{K}_D(2|1) &=& i\,\delta^4(2|1)\label{propD}
\end{eqnarray}
with $\delta^4(2|1)=\delta(t_2-t_1)\delta^3(\bm x_2-\bm x_1)$. Let
us finally remark that these relativistic quantum wave equations
have been recently of particular interest~\cite{delgado} also in the
field of quantum optical phenomena and of quantum information.

\section{Infinite divisibility--preserving modifications}\label{modifications}

\noindent The relativistic, time--like four--momentum $p=\left(E/c\,,\bm p\right)$ obeys the
relation (here $\bm p^2$ will represent the square modulus of the tri--vector $\bm p$)
\begin{equation}\label{pquadro}
    p^2=\frac{E^2}{c^2}-\bm p^2=m^2c^2\geq0
\end{equation}
so that the hamiltonian dependence of energy on momentum is
\begin{equation}\label{hamilt}
    E(\bm p)=\sqrt{m^2c^4+\bm p^2c^2}=mc^2\sqrt{1+\frac{\bm
    p^2}{m^2c^2}}\,.
\end{equation}
This relativistic energy $E$ containing a rest mass term $mc^2$, the
kinetic energy $E_0$  in a dimensionless form becomes
\begin{equation}\label{Tdimless}
    \frac{E_0(\bm p)}{mc^2}=\sqrt{1+\frac{\bm
    p^2}{m^2c^2}}-1.
\end{equation}
By taking now
\begin{equation*}
    \eta=-\,\frac{E_0}{mc^2},\qquad\bm u=\frac{\bm p}{amc}
\end{equation*}
where $a$ is a constant with the dimensions of a length, $\eta$ will
be dimensionless while $\bm u$ will be the reciprocal of a length,
and the equation\refeq{Tdimless} becomes
\begin{equation*}
    \eta(\bm u)=1-\sqrt{1+a^2\bm u^2}
\end{equation*}
namely -- not surprisingly -- it takes the three-dimensional form of the logarithmic
characteristic giving rise to the relativistic quantum equations in the Section~\ref{rqm}.

Our purpose consists now in proposing a modification of $\eta(\bm u)$ that preserves the infinite
divisibility of the law, and eventually produces changes in the forceless equations of motion for
a particle -- both at the classical and at the quantum level -- in comparison with the classical
and quantum motions given by the equations\refeq{KG} and\refeq{D}. To this end we
modify\refeq{hamilt} in the following way
\begin{equation}\label{etotmod}
    E(\bm p)=mc^2\sqrt{1+\frac{\bm
    p^2}{m^2c^2}+f\left(\frac{p^2}{m^2c^2}\right)}
\end{equation}
where $f$ is a -- possibly small -- dimensionless, smooth function
of the relativistic scalar $p^2/m^2c^2$. Of course this modification
entails that $p^2$ no longer coincides with $m^2c^2$ since the
standard relation\refeq{pquadro} is now changed into
\begin{equation}\label{pquadromod}
    p^2=\frac{E^2}{c^2}-\bm p^2=m^2c^2+m^2c^2f\left(\frac{p^2}{m^2c^2}\right).
\end{equation}
As we will see in the following this also implies that the mass no longer is $m$: it will take
instead one or more values depending on the choice of $f$. In fact it could appear to be
preposterous to introduce a function $f$ of an argument which after all is a constant (albeit
different from 1). However we will show that this will lend us the possibility of having both a
mass spectrum, and a new wave equation when -- in the next section -- we will quantize our
classical relations. Moreover it will be argued in the following that in this way the
corresponding modified logarithmic characteristic $\eta$ will remain infinitely divisible: a
feature that is instrumental for a connection to a suitable underlying L\évy process.

To see that we first remark that\refeq{pquadromod} defines the total particle energy $E$ in an
implicit form. To find it explicitly we rewrite\refeq{pquadromod} in a dimensionless form as
\begin{equation*}
    \frac{p^2}{m^2c^2}=1+f\left(\frac{p^2}{m^2c^2}\right),
\end{equation*}
and then, by taking $g(x)=x-f(x)$, we just observe that the former equation requires that $x$ be
solution of $g(x)=1$, namely
\begin{equation*}
    g\left(\frac{p^2}{m^2c^2}\right)=\frac{p^2}{m^2c^2}-f\left(\frac{p^2}{m^2c^2}\right)=1.
\end{equation*}
If then $g^{-1}(1)$ represents one of the (possibly many) solutions of this equation, we could
write
\begin{equation*}
    \frac{p^2}{m^2c^2}=g^{-1}(1)
\end{equation*}
so that we have
\begin{equation*}
    p^2=\frac{E^2}{c^2}-\bm p^2=m^2c^2g^{-1}(1)
\end{equation*}
which can be interpreted as a simple mass re-scaling from $m$ to one of the (possibly many) values
$M=m\sqrt{g^{-1}(1)}$. The new hamiltonian then is
\begin{equation}\label{newhamilt}
    E(\bm p)=\sqrt{m^2c^4g^{-1}(1)+\bm p^2c^2}=Mc^2\sqrt{1+\frac{\bm
    p^2}{M^2c^2}}
\end{equation}
and its kinetic part (by applying the same re-scaling also to the subtracted rest mass term) is
\begin{eqnarray*}
    E_0(\bm p)&=&E(\bm p)-mc^2\sqrt{g^{-1}(1)}\\
    &=&Mc^2\sqrt{1+\frac{\bm
    p^2}{M^2c^2}}-Mc^2.
\end{eqnarray*}
Hence the main consequence of our modification consists of a
re-scaling of the mass value ($m\to M$) at a purely classical level.
This fact is apparently helpful because it is straightforward to see
now that the new associated logarithmic characteristic $\eta$ is
again infinitely divisible, and hence still produces acceptable
L\évy processes. But there is more: since $g^{-1}(1)$ can take
several different real and positive values, by means of our
modification\refeq{etotmod} we have introduced an entire mass
spectrum: indeed in the rest frame of the particle we have
\begin{equation}\label{mass}
M=E_{cm}/c^2=m\sqrt{g^{-1}(1)}
\end{equation}

\section{Quantum equations of motion}


It is important to remark that while the equation\refeq{etotmod}
allows a peculiar transition to quantum mechanics ($E\to
i\hbar\partial_t$, $\bm p\to-i\hbar\bm\nabla$) if we interpret this
energy formula as a new hamiltonian operator, namely it leads to
\begin{equation}\label{releq3mod}
    i\hbar\partial_t\psi(x,t)= mc^2\sqrt{1-\frac{\hbar^2}{m^2c^2}\bm\nabla^2+f\left(\frac{\Box}{m^2c^2}\right)}\,\psi(x,t)
\end{equation}
the equation\refeq{newhamilt} gives instead the usual Klein--Gordon equation\refeq{releq3} with
just a possibly re-scaled mass $M=m\sqrt{g^{-1}(1)}$. In fact from\refeq{releq3mod}  one obtains
now a \textit{modified} Klein--Gordon equation for both the wave function $\psi$ and its
corresponding propagator $\mathcal{K}_{KG}(2|1)$ (from here on $\hbar=c=1$)
\begin{eqnarray}
  \left[\square-m^2f\left(\frac{1}{m^2}\,\square\right)-m^2\right]\psi &=& 0,\\
  \left[\square_2-m^2f\left(\frac{1}{m^2}\,\square_2\right)-m^2\right]\mathcal{K}_{KG}(2|1)\label{newpropKG} \\
  &=& \delta^4(2|1)\nonumber
\end{eqnarray}
and by standard methods~\cite{bjorken} the \emph{modified} Dirac
spinor equations
\begin{eqnarray}
  \left[i\gamma_\mu\partial^\mu-m\sqrt{1+f\left(\frac{1}{m^2}\,\square\right)}\,\right]\psi  &=& 0\\
  \left[i\gamma_\mu\partial_2^\mu-m\sqrt{1+f\left(\frac{1}{m^2}\,\square_2\right)}\,\right]\mathcal{K}_D(2|1)\label{newpropD}\\
  &=& i\delta^4(2|1)\nonumber
\end{eqnarray}
In the momentum space (with Fourier transforms in four dimensions) these equations become much
simpler: more precisely we have
\begin{eqnarray*}
  \mathcal{\mathcal{K}}_{KG}(p^2) &=& \frac{1}{p^2-m^2\left[1+f(p^2/m^2)\right]+i\epsilon} \\
  \mathcal{\mathcal{K}}_D(p^2) &=& \frac{1}{\gamma^\mu p_\mu-m\sqrt{1+f(p^2/m^2)}+i\epsilon}
\end{eqnarray*}
We notice that $\mathcal{K}_D(2|1)$ is in our case simply related to the $\mathcal{K}_{KG}(2|1)$
(like in the usual case) as
\begin{equation*}
  \mathcal{K}_D(2|1)=i\left(i\dslash_2+m\sqrt{1+f(\square_2/m^2)}\right)\mathcal{K}_{KG}(2|1)
\end{equation*}

\section{Phenomenology: quark and lepton masses}

The equations\refeq{newpropKG} and\refeq{newpropD} generalize the well known propagator
equations\refeq{propKG} and\refeq{propD} which derive from QED and QCD at zero order (in absence
of interaction terms). For future developments we recall that the Lagrangian density of QCD
is\footnote{Here $g_s$ is the QCD coupling constant, $T^a_{ij}$ and $f_{abc}$ are the $SU(3)$
color matrices and structure constants respectively, and $A_\mu^a$ the eight Yang--Mills gluon
fields; $\psi_i^q$ are the Dirac 4-spinors associated with each quark field of color $i$ and
flavor $q$.}, up to gauge fixing terms:
\begin{equation*}
    \mathcal{L}=-\frac{1}{4}F_{\mu\nu}^aF_a^{\mu\nu}+\sum_q\overline{\psi}_i^{\,q}[i\gamma^\mu(D_\mu)_{ij}-m_q\delta_{ij}]\psi_j^q
\end{equation*}
where $F_{\mu\nu}^a = \partial_\mu A^a_\nu-\partial_\nu A^a_\mu+g_sf_{abc}A^b_\mu A^c_\nu$, and
the insertion of interaction terms is done with the minimal interaction by substituting the simple
derivative $\partial_\mu$ with the covariant one $D_\mu$ where we have respectively for QED and
QCD
\begin{eqnarray*}
    D_\mu&\equiv&\partial_\mu-ieA_\mu \\
    \left(D_\mu\right)_{ij}&\equiv&\delta_{ij}\partial_\mu-ig_sT^a_{ij}A^a_\mu.
\end{eqnarray*}
The Standard Model (SM) $SU_c(3)\times SU_L(2)\times U(1)$ treats both strong and electro--weak
interactions: within this scheme the modified $\eta(\bm u)$ leads to new interesting consequences.
We begin by considering the Feynman rules in perturbation theory in presence of the modified zero
order propagator for both spin $\frac{1}{2}$ (quarks and leptons) and spin $1$ (gluons, vector
weak interacting Bosons). The amplitude $A$ for a fermion that propagates from vertex $X$ to
vertex $Y$ if expanded looks as follows: $A=A^{(0)}+A^{(1)}+A^{(2)}+\ldots$ The lowest order is
\begin{equation*}
    A^{(0)}=Y\frac{i}{\gamma^\mu p_\mu-m\sqrt{1+f(p^2/m^2)}+i\epsilon}X.
\end{equation*}
It is then possible that the Fermion emits and reabsorbs a virtual vector boson from $X$ to $Y$:
\begin{eqnarray*}
    A^{(1)}&=&4\pi g_s^2Y\int d^4k\,\frac{\gamma^\mu}{\gamma^\rho p_\rho-m\sqrt{1+f(p^2/m^2)}}\,\frac{1}{(p-k)^2}\\
         &&\qquad\qquad\times\frac{1}{k^\nu\gamma_\nu-m\sqrt{1+f(k^2/m^2)}+i\epsilon}\\
         &&\qquad\qquad\qquad\times\frac{\gamma_\mu}{\gamma^\rho p_\rho-m\sqrt{1+f(p^2/m^2)}}\,X
\end{eqnarray*}
We choose now $f(x)$ in such a way that it makes finite the integral
\begin{equation}\label{integral}
    C=\gamma^\mu\int\frac{d^{\,4}k}{\gamma^\rho k_\rho-m\sqrt{1+f\left(k^2/m^2\right)}+i\epsilon}\,\frac{1}{(p-k)^2}\,\gamma_\mu
\end{equation}
One may notice that $f(x)$ behaves as a smooth \emph{cut-off} in a procedure of Regularization at
each order in QCD (and QED). The integral $C$ is an invariant of the form $C=A(p^2)\pslash-B(p^2)$
and its integrand is also present as a factor in higher order terms, thus producing convergence.
In a similar way one expects that the representation of the complete fermionic propagator, as well
as its zero-order, is made up of two additive terms, in momentum space, each of them exhibiting
simple analyticity properties except for a limited number of poles and branch
points\footnote{Likewise one can reason for the complete and zero order propagator of basic bosons
(gluons, $W^\pm,Z^0$, Higgs).}~\cite{cutkosky}: for the zero order we have
\begin{equation*}
    \frac{i}{\pslash-m\sqrt{1+f(p^2/m^2)}}=i\,\frac{\pslash+m\sqrt{1+f(p^2/m^2)}}{p^2-m^2[1+f(p^2/m^2)]}
\end{equation*}
Let us now reconsider the equation\refeq{integral}. The simplest expression for $f(x)$ compatible
with $C$ finite is a polynomial of third degree in $x$:
\begin{equation}\label{polynomial}
    f(x)=\lambda_1x+\lambda_2x^2+\lambda_3x^3
\end{equation}
and this is suggestively connected with the possibility of having a mass spectrum. Indeed, as
stated in the Section~\ref{modifications}, the spectrum is produced by the multiple solutions of
the equation $g(x)=x-f(x)=1$, and we achieve three values that, under proper conditions, might
correspond to three masses. If we consider the three (real and positive) zeros $x_1, x_2$ and
$x_3$ of the polynomial  $g(x)-1=x-f(x)-1$ we easily find the following simple algebraic relations
with the $\lambda$'s:
\begin{eqnarray*}
    \lambda_1&=&1-\left(\frac{1}{x_1}+\frac{1}{x_2}+\frac{1}{x_3}\right)\\
    \lambda_2&=&\frac{1}{x_1x_2}+\frac{1}{x_1x_3}+\frac{1}{x_2x_3}\,,\qquad
    \lambda_3=\frac{-1}{x_1x_2x_3}
\end{eqnarray*}
The connections with the possible experimental physical masses are $M_1=m\sqrt{x_1},\,
M_2=m\sqrt{x_2},\, M_3=m\sqrt{x_3}$. If the three poles in the free (zero order) propagator are
real and positive (with proper residues), with appropriate values of the $\lambda$'s, they allow
the interpretation of physical basic masses of fermions (quark or leptons) belonging to the three
different families of the Standard Model. To be more specific we get two different propagators for
quarks, one with charge $-\frac{1}{3}$ ($d,s,b$ quarks) and another with charge $+\frac{2}{3}$
($u,c,t$ quarks). Similarly for charged leptons (charge $-1$ and spin $\frac{1}{2}$) we get one
propagator.

\begin{table}
\centering
\begin{tabular}{|c|c|c|}
 \hline
 $m_d$ & $m_s$ & $m_b$ \\
 \hline\hline
 $3\times10^{-3}$ & $70\times10^{-3}$ & $4.13$  \\
 \hline
 $7\times10^{-3}$ & $120\times10^{-3}$ & $4.27$  \\
 \hline\hline
 $\lambda_1$ & $\lambda_2$ & $\lambda_3$ \\
 \hline\hline
 $-1.84\times10^{-3}$ & $1.84\times10^{-3}$ & $-9.69\times10^{-10}$\\
 \hline
 $-3.41\times10^{-3}$ & $3.41\times10^{-3}$ & $-9.14\times10^{-9}$\\
 \hline
\end{tabular}
\caption{Estimated values of the $\lambda$'s in\refeq{polynomial}
for quarks with charge $-\frac{1}{3}$. Masses are in
Gev/c$^2$.}\label{lambdaval1}
\vspace{10pt} \centering
\begin{tabular}{|c|c|c|}
 \hline
 $m_u$ & $m_c$ & $m_t$ \\
 \hline\hline
 $1.5\times10^{-3}$ & $1.16$ & $171.2$  \\
 \hline
 $3.0\times10^{-3}$ & $1.34$ & $174.0$  \\
 \hline\hline
 $\lambda_1$ & $\lambda_2$ & $\lambda_3$ \\
 \hline\hline
 $-1.67\times10^{-6}$ & $1.67\times10^{-6}$ & $-1.28\times10^{-16}$\\
 \hline
 $-5.01\times10^{-6}$ & $5.01\times10^{-6}$ & $-1.49\times10^{-15}$\\
 \hline
\end{tabular}
\caption{Estimated values of the $\lambda$'s in\refeq{polynomial}
for quarks with charge $\frac{2}{3}$. Masses are in
Gev/c$^2$.}\label{lambdaval2}
\vspace{10pt} \centering
\begin{tabular}{|c|c|c|}
 \hline
 $m_e$ & $m_\mu$ & $m_\tau$  \\
 \hline\hline
 $5.11\times10^{-4}$ & $105.6\times10^{-3}$ & $1.77$  \\
 \hline\hline
 $\lambda_1$ & $\lambda_2$ & $\lambda_3$\\
 \hline\hline
 $-2.35\times10^{-5}$ & $2.35\times10^{-5}$ & $-1.95\times10^{-12}$\\
 \hline
\end{tabular}
\caption{Estimated values of the $\lambda$'s in\refeq{polynomial}
for leptons with charge $-1$. Masses are in
Gev/c$^2$.}\label{lambdaval3}
 \end{table}

In the Tables~\ref{lambdaval1},~\ref{lambdaval2} and~\ref{lambdaval3} we give a few examples of
numerical values of the $\lambda$ parameters in equation\refeq{polynomial} for quark and lepton
masses (in Gev/c$^2$) taken from the Particle Data Group~\cite{PDG}. For the quarks with charge
$-\frac{1}{3}$ and different mass estimates we obtain the results listed in the
Table~\ref{lambdaval1}, while for quarks with charge $\frac{2}{3}$ we get the results of the
Table~\ref{lambdaval2}, and finally for charged leptons with charge $-1$ we have the results of
Table~\ref{lambdaval3}. According to our model a significant contribution to the fermion mass
spectrum derives from the poles of the zero order propagator, whereas the role of the interaction
terms might become complementary and can be estimated in (renormalized) perturbation theory.

Our $\lambda$'s also produce the regularization at each order in QCD
(and QED) perturbation theory. Furthermore we know that after
regularization  the approximate representation of the propagator
tends to a finite limit (exact propagator) due to the
renormalization mechanism. More precisely in these field theories
the calculated renormalized physical quantities are supposed to
become independent of the \emph{cut-off}. The latter must disappear
in the transition from regularization to renormalization.
Consequently the $\lambda$'s remain in a finite fixed number and
tend to definite (real) values as the \emph{cut-off} cancels out.
Within this scenario the poles representing physical masses (3 in
our case) remain stable (even if shifted partly with respect to
those computed approximately by us), because of to the assumed
analytical properties of the renormalized
propagator~\cite{feynman,cutkosky}.

\section{Conclusions}

We have proposed a modification of the classical relativistic hamiltonian that allows the presence
of several masses without changing its basic structure. This modification does not affect the
infinite divisibility of the laws that are at the basis of the correspondence between stochastic
processes and L\'evy--quantum mechanics equations. However we discovered that the mentioned
modification suggests a reformulation of the relativistic equations for wave functions and
propagators in such a way that a suitable choice of the background noise  produces a convergence
in the perturbative contributions. To this purpose  we remarked that a modification -- with
respect to the one given by equation\refeq{relEp} -- of the logarithmic characteristic $\eta(\bm
u)$ by the insertion of the cut-off $f(x)$ allows to proceed to Regularization first, and then
Renormalization of the two-point function of QCD. There are three parameters in our
phenomenological $f(x)$ which is a third degree polynomial; the latter appears as the simplest
choice that produces convergence in the integrals representing high order contributions to the
fermion and boson propagators. Such parameters create three different poles in the zero-order
propagators and allow the interpretation of a physical system with three different masses under
precise constraints on $f(x)$. The masses might be related to the three families of the Standard
Model. We like to point out that from the analyticity properties of the renormalized propagators,
the mentioned poles tend to stabilize in the limit toward the complete solution (even if shifted
with respect to the zero order ones) whereas the cut-off is expected to disappear because of the
regular renormalizable QCD theory~\cite{feynman}.

\end{document}